\documentclass[aps,prc,preprint,superscriptaddress,showpacs]{revtex4}
\usepackage[dvips]{graphicx}
\usepackage{amsmath,amssymb}
\usepackage{ulem}

\newcommand{\beq}{\begin{equation}}
\newcommand{\eeq}{\end{equation}}
\newcommand{\bea}{\begin{eqnarray}}
\newcommand{\eea}{\end{eqnarray}}

\DeclareSymbolFont{boldletters}{OML}{cmm} {b}{it}
\DeclareSymbolFontAlphabet{\mathbit}{boldletters}
\DeclareMathSymbol{\alpha}{\mathalpha}{letters}{"0B}
\DeclareMathSymbol{\beta}{\mathalpha}{letters}{"0C}
\DeclareMathSymbol{\gamma}{\mathalpha}{letters}{"0D}
\DeclareMathSymbol{\delta}{\mathalpha}{letters}{"0E}
\DeclareMathSymbol{\epsilon}{\mathalpha}{letters}{"0F}
\DeclareMathSymbol{\zeta}{\mathalpha}{letters}{"10}
\DeclareMathSymbol{\eta}{\mathalpha}{letters}{"11}
\DeclareMathSymbol{\theta}{\mathalpha}{letters}{"12}
\DeclareMathSymbol{\iota}{\mathalpha}{letters}{"13}
\DeclareMathSymbol{\kappa}{\mathalpha}{letters}{"14}
\DeclareMathSymbol{\lambda}{\mathalpha}{letters}{"15}
\DeclareMathSymbol{\mu}{\mathalpha}{letters}{"16}
\DeclareMathSymbol{\nu}{\mathalpha}{letters}{"17}
\DeclareMathSymbol{\xi}{\mathalpha}{letters}{"18}
\DeclareMathSymbol{\pi}{\mathalpha}{letters}{"19}
\DeclareMathSymbol{\rho}{\mathalpha}{letters}{"1A}
\DeclareMathSymbol{\sigma}{\mathalpha}{letters}{"1B}
\DeclareMathSymbol{\tau}{\mathalpha}{letters}{"1C}
\DeclareMathSymbol{\upsilon}{\mathalpha}{letters}{"1D}
\DeclareMathSymbol{\phi}{\mathalpha}{letters}{"1E}
\DeclareMathSymbol{\chi}{\mathalpha}{letters}{"1F}
\DeclareMathSymbol{\psi}{\mathalpha}{letters}{"20}
\DeclareMathSymbol{\omega}{\mathalpha}{letters}{"21}
\DeclareMathSymbol{\varepsilon}{\mathalpha}{letters}{"22}
\DeclareMathSymbol{\vartheta}{\mathalpha}{letters}{"23}
\DeclareMathSymbol{\varpi}{\mathalpha}{letters}{"24}
\DeclareMathSymbol{\varrho}{\mathalpha}{letters}{"25}
\DeclareMathSymbol{\varsigma}{\mathalpha}{letters}{"26}
\DeclareMathSymbol{\varphi}{\mathalpha}{letters}{"27}
\DeclareMathSymbol{\Gamma}{\mathalpha}{letters}{"00}
\DeclareMathSymbol{\Delta}{\mathalpha}{letters}{"01}
\DeclareMathSymbol{\Theta}{\mathalpha}{letters}{"02}
\DeclareMathSymbol{\Lambda}{\mathalpha}{letters}{"03}
\DeclareMathSymbol{\Xi}{\mathalpha}{letters}{"04}
\DeclareMathSymbol{\Pi}{\mathalpha}{letters}{"05}
\DeclareMathSymbol{\Sigma}{\mathalpha}{letters}{"06}
\DeclareMathSymbol{\Upsilon}{\mathalpha}{letters}{"07}
\DeclareMathSymbol{\Phi}{\mathalpha}{letters}{"08}
\DeclareMathSymbol{\Psi}{\mathalpha}{letters}{"09}
\DeclareMathSymbol{\Omega}{\mathalpha}{letters}{"0A}

\def\delsla{\!\not\!\partial}
\begin{document}
\preprint{SAGA-HE-229-06}
\title{Chiral phase transition in an extended NJL model 
with higher-order multi-quark interactions}

\author{Kouji Kashiwa}
\affiliation{Department of Physics, Graduate School of Sciences, Kyushu University,
             Fukuoka 812-8581, Japan}

\author{Hiroaki Kouno}
\email[]{kounoh@cc.saga-u.ac.jp}
\affiliation{Department of Physics, Saga University,
             Saga 840-8502, Japan}

\author{Tomohiko Sakaguchi}
\affiliation{Department of Physics, Graduate School of Sciences, Kyushu University,
             Fukuoka 812-8581, Japan}

\author{Masayuki Matsuzaki}
%\email[]{matsuza@fukuoka-edu.ac.jp}
\affiliation{Department of Physics, Fukuoka University of Education, 
             Munakata, Fukuoka 811-4192, Japan}

\author{Masanobu Yahiro}
%\email[]{yahiro2scp@mbox.nc.kyushu-u.ac.jp}
\affiliation{Department of Physics, Graduate School of Sciences, Kyushu University,
             Fukuoka 812-8581, Japan}

\date{\today}

\begin{abstract}
The chiral phase transition is studied in 
an extended Nambu--Jona-Lasinio model with eight-quark interactions. 
Equations for scalar and vector quark densities, derived in 
the mean field approximation, are nonlinear and mutually coupled. 
The scalar-type nonlinear term hastens the restoration 
of chiral symmetry, while the scalar-vector mixing term makes 
the transition sharper. 
The scalar-type nonlinear term shifts the critical endpoint 
toward the values predicted by lattice QCD simulations and the QCD-like theory. 
\end{abstract}

\maketitle

%%%%%%%%%%%%%%%%%%%%%%%%%%%%%%%%%%%%%%%%%%%%%%%%%%%%%%%%%%%%%%%%%%%%%%%%%%%
%%%%%  Introduction 
%%%%%%%%%%%%%%%%%%%%%%%%%%%%%%%%%%%%%%%%%%%%%%%%%%%%%%%%%%%%%%%%%%%%%%%%%%%

Qualitative properties in quantum chromodynamics (QCD) 
at high temperatures and densities attract much attention. 
One of the most important recent findings is strong correlations 
in the quark gluon plasma just above the critical temperature; 
it is realized as the near perfect fluidity~\cite{Lee,GM,Shu} and 
the mesonic correlations~\cite{AH,Gho}. 

With the aid of the progress in computer power, lattice QCD simulations have become feasible for thermal systems at zero or small density~\cite{Kog,ZF}. 
At high density, however, lattice QCD is still not feasible due to 
the sign problem. As an approach complementary to the first-principle 
lattice simulations, we can consider several effective models. 
One of them is the Nambu--Jona-Lasinio (NJL) model. 
Since it was proposed~\cite{NJ1}, although this is a model of chiral symmetry that does not possess a confinement mechanism, this model has been widely 
used~\cite{Kle,HK1} in the mean field approximation (MFA), for example, for 
analyses of the critical endpoint of chiral transition~\cite{AY,Buballa,BR,Sca,Fuj,KKKN}. 

Although the NJL model is recognized as an useful method for understanding 
the chiral symmetry breaking, only a few studies were done so far 
on roles of higher-order multi-quark interactions~\cite{Osipov1,Osipov2}, 
except for the case of the six-quark interaction coming 
from the 't Hooft determinant interaction~\cite{tHooft}. 
The NJL model is an effective theory of QCD, so there is 
no reason, in principle, 
why higher-order multi-quark interactions are excluded. 
In the nonperturbative renormalization group method~\cite{AMSTT2}, 
such higher-order interactions are produced 
as a result of quantum effects in the high momentum region. 
Such effects should be included in the low-energy effective action 
from the beginning. 
Thus, it is quite meaningful 
to study effects of higher-order interactions on the 
chiral phase transition.

In this paper, we consider an extended NJL model that 
newly includes eight-quark interactions and analyze 
roles of such higher-order interactions on 
the chiral phase transition. % by using the mean field approximation. 
It is well known that the original NJL model predicts 
a critical endpoint to appear at 
a lower temperature $(T)$ and a higher chemical potential ($\mu$) 
than lattice QCD~\cite{ZF} and 
the QCD-like theory~\cite{KMT,HTF} do. 
We will show in this paper that a scalar-type eight-quark interaction 
shifts the critical endpoint toward values 
predicted by the lattice simulations and the QCD-like theory.

As for the repulsive vector-type four-quark interaction 
$(\bar{q}\gamma_\mu q)^2$, 
it is well-known that it makes the chiral phase transition weaker 
in the low $T$ and high $\mu$ region, so there is a possibility that 
the transition becomes a crossover in the region when the interaction 
is strong enough~\cite{Buballa,KKKN}.
In this point of view, an $absence$ of 
the vector-type four-quark interaction 
may be preferable in the high density region; 
the absence is also 
supported by works of Refs. \cite{Bentz} and \cite{Detmold}. 

On the contrary, in the relativistic meson-nucleon theory~\cite{Walecka}, 
the repulsive force mediated by vector mesons 
is necessary to account for the saturation property of nuclear matter. 
Using the auxiliary field method, we can convert quark-quark interactions 
to meson-quark interactions; 
for example, see Refs.~\cite{KS,Sakaetal,KSKHTMY} and references therein. 
It is then natural to think that there exists 
a relation between the meson-nucleon interaction and 
the quark-quark interaction in the NJL model, 
since a nucleon is composed of three constituent quarks each of which has 
a large effective mass as a result of the spontaneous symmetry breaking (SSB) 
of chiral symmetry. 
In this point of view, the vector-type four-quark interaction 
$(\bar{q}\gamma_\mu q)^2$ is $indispensable$ 
around $\mu \sim 308$~MeV corresponding to the normal nuclear density region. 
Thus, it is expected that the vector-type interaction is sizable 
in the normal density region but suppressed in the higher density region.

%This mechanism is considered to be realized by introducing 
%higher-order multi-quark interactions. 
In the relativistic meson-nucleon theory, it is known that 
nonlinear meson-nucleon interactions suppress 
the effective coupling between mesons and 
nucleons in the higher density region~\cite{TKHN,KHT,Tetal,KSKHTMY}. 
It is then strongly expected that a similar situation takes place in the NJL model 
as soon as higher-order multi-quark interactions are introduced. 
This is shown in this paper.

This is the first work to study roles of eight-quark interactions 
on phase transitions, so we concentrate our analysis 
on the chiral phase transition and use a simple model with two flavor quarks. 
Effects of the higher-order interactions on the color superconductivity 
will be discussed in a forthcoming paper. 

%%%%%%%%%%%%%%%%%%%%%%%%%%%%%%%%%%%%%%%%%%%%%%%%%%%%%%%%%%%%%%%%%%%%%%%%%%%%%%%%%%%%%  Formalism
%%%%%%%%%%%%%%%%%%%%%%%%%%%%%%%%%%%%%%%%%%%%%%%%%%%%%%%%%%%%%%%%%%%%%%%%%%%%%%%%
We start with the following chiral-invariant Lagrangian density 
with two flavor quarks 
%%%%%%%%%%%%%%%%
\begin{eqnarray}
{\cal L} &=& {\bar q}( i \delsla-m_0 ) q
           + \Bigr[{g}_{2,0} \Bigl( 
             (\bar q q)^2 + (\bar q i \gamma_5  \vec{\tau} q)^2 \Bigl)
           +{g}_{4,0}\Bigl( 
             (\bar q q)^2 + ( \bar q i\gamma_5  \vec{\tau} q)^2 \Bigr)^2 
                                                          \nonumber \\
         &&-{g_{0,2}}( \bar q \gamma^\mu q)^2
           -{g_{2,2}} \Bigl( (\bar q q)^2  
           +( \bar q i\gamma_5  \vec{\tau} q)^2 \Bigr) (\bar q \gamma^\mu q)^2
                 + \cdots \Bigr], 
\label{eq:E1}
\end{eqnarray}
%%%%%%%%%%%%%%
where $q$, $m_0$ and 
$g_{i,j}$
stand for a quark field, a bare quark mass and 
coupling constants. 
Here, we consider only four- and eight-quark interactions by ignoring  
the higher-order interactions denoted by dots in Eq. (\ref{eq:E1}). 
Furthermore, we disregard interactions including isovector-vector current 
not important in symmetric quark matter, and also does 
the vector-type eight-quark interaction $(\bar{q}\gamma_\mu q)^4$, 
because the expectation value of the vector current $\bar{q}\gamma_0 q$ 
is smaller than that of the scalar one $\bar{q}q$, unless 
the chemical potential is quite large.
The mean field approximation reduces the Lagrangian to 
%%%%%%%%%%%%%%
\begin{eqnarray}
{\cal L}_{\rm MFA} 
&=& {\bar q}( i \delsla-m_0 ) q
\nonumber\\
&&+\left[2{g}_{2,0}
+4{g}_{4,0}\Bigl( \langle {\bar q}q \rangle^2 
 + \langle {\bar q} i\gamma_5 \vec{\tau} q \rangle ^2\Bigr)
-2{g}_{2,2}\langle {\bar q}\gamma^\mu q \rangle^2\right]
\Bigl(\langle {\bar q}q \rangle {\bar q}q +
\langle {\bar q} i\gamma_5 \vec{\tau} q \rangle
{\bar q}i\gamma_5 \vec{\tau}q   \Bigr)
\nonumber\\
&&-\left[2{g}_{0,2}
+2{g}_{2,2}\Bigl( \langle {\bar q}q \rangle^2 
 + \langle {\bar q} i\gamma_5 \vec{\tau} q \rangle ^2\Bigr)\right]
\langle {\bar q}\gamma^\mu q \rangle {\bar q}\gamma_\mu q
\nonumber\\
&&  -{g}_{2,0} \left( \langle {\bar q}q \rangle^2 +\langle {\bar q}i\gamma_5  \vec{\tau} q \rangle^2\right) 
-3{g}_{4,0} \left( \langle {\bar q}q \rangle^2+\langle {\bar q}i\gamma_5  \vec{\tau} q \rangle^2\right)^2
\nonumber\\
&&+3{g}_{2,2} \left( \langle {\bar q}q \rangle^2
+\langle {\bar q}i\gamma_5  \vec{\tau} q \rangle^2\right) 
\langle {\bar q}\gamma^\mu q \rangle^2
+{g}_{0,2} \langle {\bar q}\gamma^\mu q \rangle^2. 
\label{eq:E5}
\end{eqnarray}
%%%%%%%%%%%%%%%
Below, just for simplicity of the notation, we will omit terms including 
the pseudo-scalar current $\bar{q}i\gamma_5\vec{\tau} q$ and 
the spatial components $\bar{q}\gamma_iq~~(i=1,2,3)$ of the vector current, 
since their expectation values vanish. 
It is convenient to introduce two auxiliary mean fields as
%%%%%%%%%%%%%
\begin{equation}
\sigma  \equiv \langle \bar{q}q \rangle ,~~~~~~~~\omega \equiv \langle \bar{q}\gamma_0q \rangle .
\label{eq:E6}
\end{equation} 
%%%%%%%%%%%%%
Using these auxiliary fields, one can rewrite ${\cal L}_{\rm MFA}$ as 
%%%%%%%%%%%%%%%%%
\begin{eqnarray}
{\cal L}_{\rm MFA}&=&{\bar q} [i \delsla -(m_0+\Sigma_{\rm s}) + \Sigma_{\rm v} \gamma_0]q
                -U \;,
\end{eqnarray}
where
\begin{eqnarray}
\Sigma_{\rm s}  &=& -\Bigl( 2{g}_{2,0} \sigma
                   +4{g}_{4,0} \sigma^3 
		   -2{g}_{2,2} \sigma \omega^2 \Bigr), ~~~~~
\Sigma_{\rm v}  = -\Bigl( 2{g}_{0,2} \omega 
		   +2{g}_{2,2} \sigma^2 \omega \Bigr), 
\label{eq:E8}
\nonumber\\
U       &=& {g}_{2,0} \sigma^2
             +3{g}_{4,0} \sigma^4
             -3{g}_{2,2} \sigma^2 \omega^2
             -{g}_{0,2} \omega^2. 
\label{eq:E9}
\end{eqnarray}
%%%%%%%%%%%%%%%%%
The thermodynamical potential $\Omega$ of the system with finite 
$T$ and $\mu$ is then obtained by 
%%%%%%%%%%%%%%%
\begin{eqnarray}
\Omega (T,\mu ) &=& -2 N_f N_c V  
                 \int \frac{d^3{\bf p}}{(2 \pi)^3} \Bigl[E_p
                 + \frac{1}{\beta} 
                    \Bigl(\log{\left(1+e^{-\beta(E_p-\mu^*)}\right)}
                    +
                          \log{\left(1+e^{-\beta(E_p+\mu^*)}\right)}\Bigr)\Big]
\nonumber\\
&&+VU, 
\label{eq:E10}
\end{eqnarray}
%%%%%%%%%%%%%%%%%%%%%%%%%%%%%%%%%%%%%%%%%%%%%%%%%%%%%%%%%%%%%%%%%%%%%%%%%%%%%%%
where $\beta =1/T$, $\mu^*=\mu+\Sigma_{\rm v}$, 
$E_p=\sqrt{{\bf p}^2+M^2}$ and 
$M$ stands for the effective quark mass defined 
as $M=m_0+\Sigma_{\rm s}$. 
The corresponding scalar and vector quark densities, 
$\rho_{\rm s}$ and $\rho_{\rm v}$, are also given by 
%%%%%%%%%%%%%%%
\begin{eqnarray}
\rho_{\rm s} &=&2N_f N_c \int \frac{d^3{\bf p}}{(2 \pi)^3}
            \frac{M}{E_p} 
            \Bigl(
                  n_q + n_{\bar q} -1
            \Bigr), ~~~~~
\rho_{\rm v} =2 N_f N_c \int \frac{d^3{\bf p}}{(2 \pi)^3}
            \Bigl(
	          n_q  - n_{\bar q}
            \Bigr) \;,~~~~~
%\nonumber\\
%n_q      &=& \frac{1}{1+\exp{\{\beta(E_p-\mu^*)\}}}, ~~~~~
%n_{\bar q}= \frac{1}{1+\exp{\{\beta(E_p+\mu^*)\}}} \;, 
\label{eq:E14}
\end{eqnarray}
where $n_q=1/[{1+\exp{\{\beta(E_p-\mu^*)\}}}]$
and $n_{\bar q}=[{1+\exp{\{\beta(E_p+\mu^*)\}}}]$. 
%%%%%%%%%%%%%%
The physical solutions of $\sigma$ and $\omega$ satisfy the stationary condition 
%%%%%%%%%%%%%%%%
\begin{eqnarray}
\left(
\begin{array}{c}
{\partial \over{\partial \sigma}}\left({\Omega\over{V }}\right) \\
{\partial  \over{\partial \omega}}\left({\Omega\over{V }}\right)
\end{array}
\right)
&=&{\cal G}^*
\left(
\begin{array}{c}
\sigma -\rho_{\rm s} \\
\rho_{\rm v}-\omega
\end{array}
\right)
=\left(
\begin{array}{c}
0 \\
0
\end{array}
\right),~~~
{\cal G}^*\equiv 
\left(
\begin{array}{cc}
G_{{\rm s}\sigma}^* & G_{{\rm v}\sigma}^* \\
G_{{\rm s}\omega}^* & G_{{\rm v}\omega}^* \\
\end{array}
\right).
\label{eq:E15}
\end{eqnarray}
%%%%%%%%%%%%%%
Here we have defined four effective couplings, 
$G_{{\rm s}\sigma}^*$, $G_{{\rm v}\sigma}^*$, $G_{{\rm s}\omega}^*$ and 
$G_{{\rm v}\omega}^*$, 
as~\cite{TKHN,KHT,Tetal,KSKHTMY}
%%%%%%%%%%%%%%%%
\begin{equation}
G_{{\rm s}\sigma}^*\equiv -{\partial \Sigma_{\rm s}\over{\partial \sigma}}
=2g_{2,0}+12g_{4,0}\sigma^2-2g_{2,2}\omega^2,~~~
G_{{\rm v}\sigma}^*=-{\partial \Sigma_{\rm v}\over{\partial \sigma}}
=4g_{2,2}\sigma\omega . 
\label{eq:scalar-G}
\end{equation}
%%%%%%%%%%%%%%
%%%%%%%%%%%%%%%%
\begin{equation}
G_{{\rm s}\omega}^*=-{\partial \Sigma_{\rm s}\over{\partial \omega}}
=-4g_{2,2}\sigma\omega=-G_{{\rm v}\sigma}^*,~~~
G_{{\rm v}\omega}^*=-{\partial \Sigma_{\rm v}\over{\partial \omega}}
=2g_{0,2}+2g_{2,2}\sigma^2. 
\label{eq:vector-G}
\end{equation}
%%%%%%%%%%%%%%
When ${\rm det}({\cal G}^*)=G_{{\rm s}\sigma}^*G_{{\rm v}\omega}^*+{G_{{\rm v}\sigma}^*}^2\neq 0$, 
which is satisfied in our analyses below, 
the matrix ${\cal G}^*$ has its inverse, and then 
the stationary condition (\ref{eq:E15}) leads to 
$\sigma =\rho_{\rm s}$ and $\omega =\rho_{\rm v}$, 
showing the consistency with Eq. (\ref{eq:E6})~\cite{KHT}. 
However, the solutions $\sigma$ and $\omega$ to the equations 
do not necessarily yield a minimum of $\Omega$.   
The solution $\sigma$ is then determined 
by searching for minima of $\Omega(\sigma,\omega(\sigma))$ in which $\omega$ is eliminated with $\omega =\rho_{\rm v}$.

Identifying  $\bar{q}i\gamma_5\vec{\tau} q$ with the pion field, 
we can define the effective coupling $G_{{\rm s}\pi}^*$ 
between pion and quark as 
%%%%%%%%%%%%%%%%
\begin{equation}
iG_{{\rm s}\pi}^*\gamma_5{\vec{\tau}}\equiv \left.{\delta^3\over{\delta q\delta \langle \bar{q}i\gamma_5\vec{\tau} q \rangle \delta \bar{q}}}{\cal L}_{\rm MFA}\right|_{\langle \bar{q}i\gamma_5\vec{\tau} q \rangle=0}=i\left(2g_{2,0}+4g_{4,0}\sigma^2-2g_{2,2}\omega^2\right)\gamma_5\vec{\tau}. 
\label{eq:E25}
\end{equation}
%%%%%%%%%%%%%%%%
In the random phase approximation (RPA), the pion mass $M_\pi$ at $T=\mu=0$ is determined with this effective coupling as 
%%%%%%%%%%%%%%%%
\begin{equation}
M_\pi^2={4m_0\over{G_{{\rm s}\pi}^*MI(M,M_\pi)}},~~~~~
I(x,y)=8N_{\rm f}N_{\rm c} \int \frac{d^3{\bf p}}{(2 \pi)^3} {1\over{\sqrt{{\bf p}^2+x^2}(4({\bf p}^2+x^2)-y^2)}}. 
\label{eq:E26}
\end{equation}
%%%%%%%%%%%%%%
Similarly, the $\sigma$-meson mass $M_\sigma$ at $T=\mu=0$ 
is determined as 
%%%%%%%%%%%%%%%%
\begin{equation}
M_\sigma^2={4m_0+32g_{4,0}\sigma^3\over{G_{{\rm s}\sigma}^*MI(M,M_\sigma)}}+4M^2.\label{eq:E28}
\end{equation}
%%%%%%%%%%%%%%
This equation indicates that $M_\sigma < 2M$ when $\sigma <0$ and $g_{4,0}>0$. 
%%%%%%%%%%%%%%%%%%%%%%%%%%%%%%%%%%%%%%%%%%%%%%%%%%%%%%%%%%%%%%%%%%%%%%%%%%%%%%%%%%%%%  Numerical Results
%%%%%%%%%%%%%%%%%%%%%%%%%%%%%%%%%%%%%%%%%%%%%%%%%%%%%%%%%%%%%%%%%%%%%%%%%%%%%%%%

Since the NJL model is nonrenormalizable, it is needed to 
introduce a cutoff in the momentum integration. 
In this paper, we use the three-dimensional momentum cutoff as 
%%%%%%%%%%%%%%%%
\begin{equation}
\int \frac{d^3{\bf p}}{(2 \pi)^3}\to 
{1\over{2\pi^2}} \int_0^\Lambda p^2dp .
\label{eq:E29}
\end{equation}
%%%%%%%%%%%%%%
Hence, the present model has six parameters, 
$m_0$, $\Lambda$, $g_{2,0}$, $g_{4,0}$, $g_{0,2}$ and $g_{2,2}$. 
We simply assume $m_0=$ 5.5 MeV. 
For the case of $g_{4,0}=0$, $\Lambda$ and $g_{2,0}$ are fixed 
to reproduce the empirical values of the pion decay constant $f_\pi=$ 93.3 MeV and the pion mass $M_\pi=$ 138 MeV. 
For the case of nonzero $g_{4,0}$, 
$\Lambda$, $g_{2,0}$ and $g_{4,0}$ are fixed 
to reproduce $f_\pi$ and $M_\pi$ above and $M_\sigma=$ 650 MeV. 
In the latter case, the contribution of the $g_{4,0}$ term to $G_{{\rm s}\sigma}^*$ is about 11 percents at $T=0$ and $\mu=0$. 
For the vector coupling constants $g_{0,2}$ and $g_{2,2}$, 
we take two extreme cases, $G_\omega =0$ and $G_\omega =G_\sigma /1.5$, 
where $G_{\sigma}\equiv G_{{\rm s}\sigma}^*|_{T=\mu =0}
=2g_{2,0}+12g_{4,0}\sigma_0^2$ and  
$G_{\omega}\equiv G_{{\rm v}\omega}^*|_{T=\mu =0}
=2g_{0,2}+2g_{2,2}\sigma_0^2$ 
for $\sigma_0$ the scalar density at $T=0$ and $\mu=0$. 
Furthermore, in order to determine each value of $g_{0,2}$ and $g_{2,2}$, we 
take two cases, $(2g_{0,2},2g_{2,2}\sigma_0^2)=(G_\omega,0)$ and  
$(2g_{0,2},2g_{2,2}\sigma_0^2)=(0.8G_\omega,0.2G_\omega)$.  
For the second case, 
%$2g_{2,2}\sigma_0^2=0.2 G_\omega$, 
$G_{{\rm v}\omega}^*$ is suppressed in the high $\mu$ region, 
as shown later. 
Table I summarizes the parameter sets we have taken, $M_\sigma$ at $T=0$ and $\mu =0$, and the type of transition at $T=0$. 
%%%%%%%%%%%%%%%%
\begin{table}[h]
\begin{center}
\begin{tabular}{lllllll}
\hline
model 
& $g_{2,0}$ & $g_{4,0} \sigma_0^2$ & $2g_{0,2}$ & $2g_{2,2} \sigma_0^2$ 
& $M_\sigma$ [GeV]~~& type 
\\
\hline
NJL
%     & 10.9960215~~~
& 5.498
& 0~~~
& 0~~~
& 0~~~
& 0.681~~~
& first order
\\
NJL + $\sigma^4$ 
%     & 10.5511154 
& 5.276
%       &0.443824292
&0.1109
& 0 
& 0
& 0.650~~~
& first order 
\\
NJL + $\omega^2$ 
& 5.498
%     & 10.9960215 
& 0 
& $ G_\omega$~~~
& 0 
& 0.681~~~
& crossover
\\
NJL + $\omega^2 + \sigma^2 \omega^2$ 
& 5.498~~~
%     & 10.9960215~~~
& 0 
& 0.8 $G_\omega$~~~
& 0.2 $G_\omega$~~~
& 0.681~~~
& crossover 
\\
NJL + $\sigma^4$ + $\omega^2$ 
& 5.276
%& 10.5511154 
%       & 0.443824292~~~
& 0.1109~~
& $ G_\omega $
& 0 
& 0.650~~~
& crossover
\\
NJL + $\sigma^4$ + $\omega^2 + \sigma^2 \omega^2$~~~
& 5.276
%     & 10.5511154 
%       & 0.443824292
& 0.1109
& 0.8 $G_\omega$
& 0.2 $G_\omega$
& 0.650~~~
& first order
\\
\hline
\end{tabular}
\caption{
Summary of the parameter sets, $M_\sigma$ at $T=0$ and $\mu =0$, 
and the type of the transition at $T=0$.  
The coupling constants are shown in 
${\rm  GeV}^{-2}$. 
The effective coupling $G_\omega$ is fixed to $7.331 ~{\rm GeV}^{-2}$ 
in the NJL+$\omega^2$ and the  NJL+$\omega^2$+$\sigma^2 \omega^2$
model and to $7.922 ~{\rm GeV}^{-2}$ 
in the NJL+$\sigma^4$+$\omega^2$ and the
 NJL+$\sigma^4$+$\omega^2$+$\sigma^2 \omega^2$  model.  
 For all cases, we take $\Lambda =0.6315$~GeV and 
 $M|_{T=\mu=0}=0.3379$~GeV and $\sigma_0=-0.03023$~GeV$^3$. 
}
\end{center}
\end{table}
%%%%%%%%%%%

Figure 1(a) shows the $T$ dependence of the effective quark mass for the case 
of $\mu=0$. 
Since the $\omega$ field has no contribution to $M$ when 
$\mu=0$, 
results are shown only for the original NJL and the NJL$+\sigma^4$ model. 
The nonlinear $\sigma^{4}$ interaction  makes 
the restoration of chiral symmetry faster, since the effective coupling, 
responsible for the SSB of chiral symmetry, becomes smaller as $T$ increases, 
as shown in Fig.1(b).  
However, the interaction keeps the transition a crossover 
as in the case of the original NJL. 
The NJL+$\sigma^4$ model yields a smaller transition temperature ($T_{\rm c}\sim$180MeV) than the original NJL model does $(T_{\rm c}\sim$190MeV). 
The value, $T_{\rm c}\sim$180MeV, is close to the one predicted by the lattice simulation~($T_{\rm c}\sim$170MeV)~\cite{ZF}. 

%%%%%%%%%%%%%%%% Fig %%%%%%%%%%%%%%%%%%%%% ->
\begin{figure}[htbp]%[H]
\begin{center}
 \includegraphics[width=7.5cm]{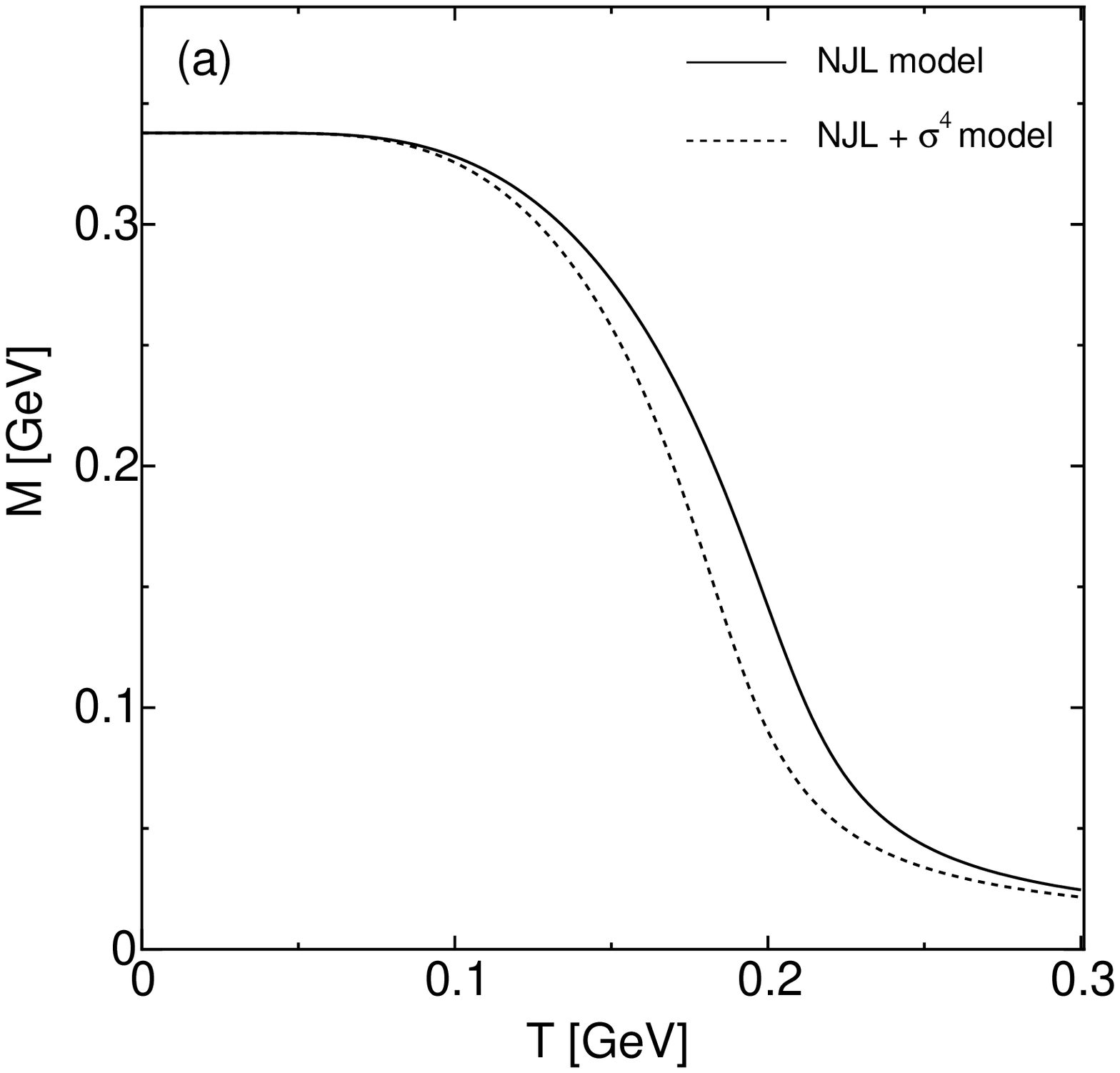} 
  \includegraphics[width=7.5cm]{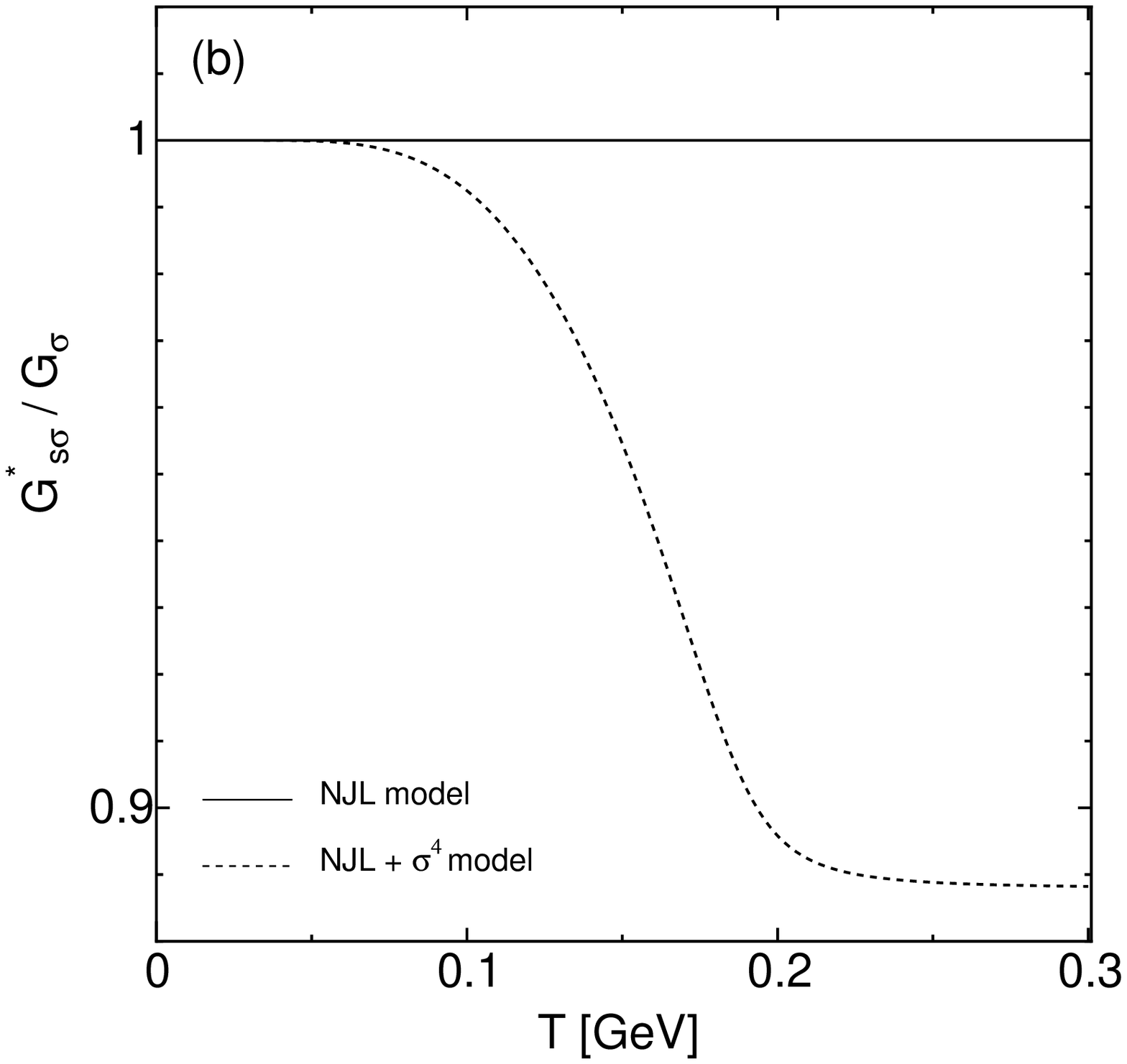} 
\end{center}
\caption{The $T$ dependence of (a) the effective quark mass and 
(b) the effective coupling in the case of $\mu =0$. 
}
\label{fig1}
\end{figure}
%%%%%%%%%%%%%%%%%%%%%%%%%%%%%%%%%%%%%%%%%%%%%%%%%

Figure 2 shows the effective quark mass as a function of $\mu$ 
with $T=0$. 
In the original NJL model, the chiral transition is a first order. 
The nonlinear $\sigma^4$ interaction makes the transition faster, as shown in 
the result of the NJL+$\sigma^4$ model.

%%%%%%%%%%%%%%%% Fig %%%%%%%%%%%%%%%%%%%%% ->
\begin{figure}[htbp]%[H]
\begin{center}
 \includegraphics[width=7.5cm]{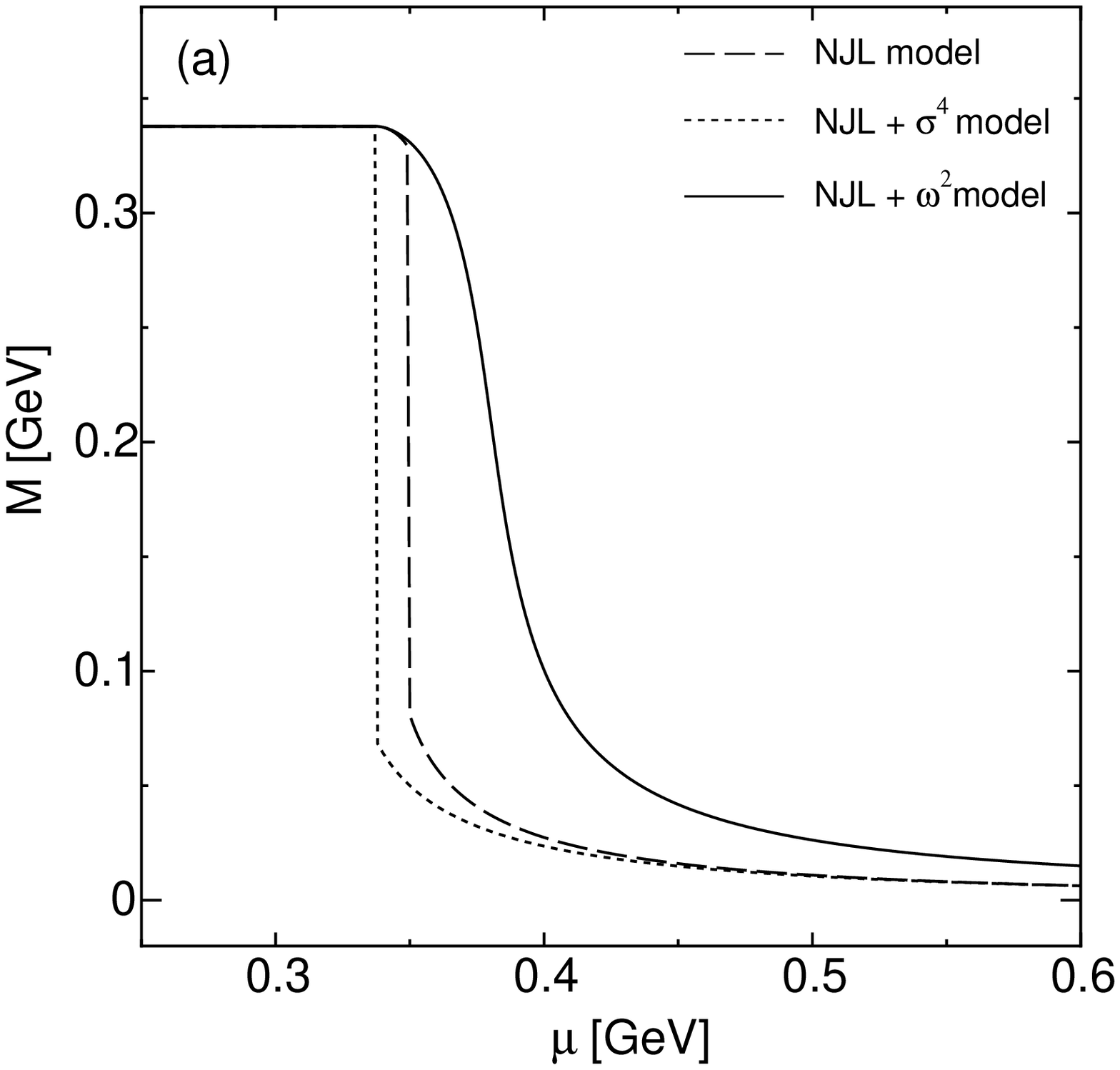} 
 \includegraphics[width=7.5cm]{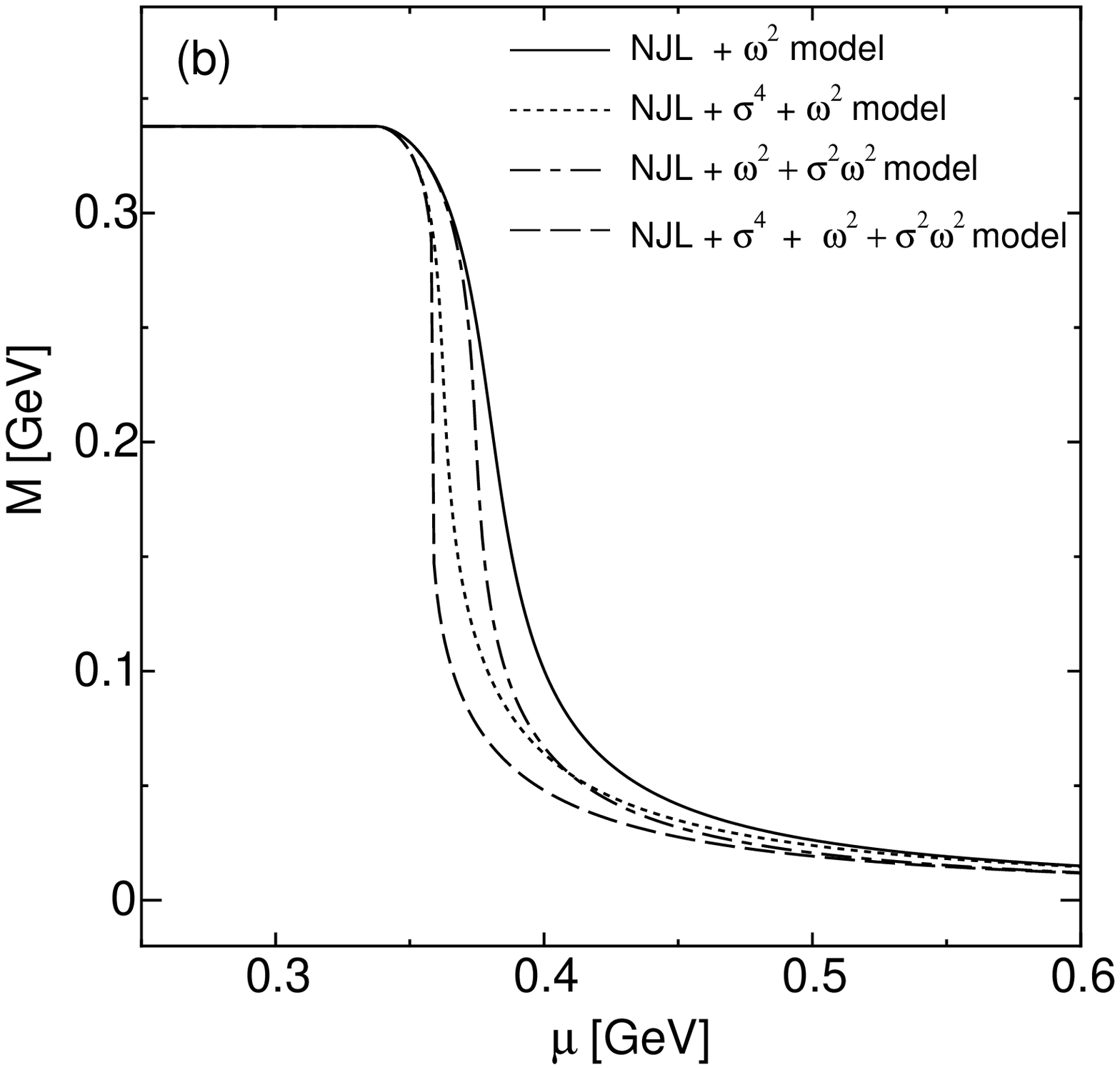}
\end{center}
\caption{The $\mu$ dependence of the effective quark mass in the case 
of $T=0$. 
}
\label{fig2}
\end{figure}
%%%%%%%%%%%%%%%%%%%%%%%%%%%%%%%%%%%%%%%%%%%%%%%%%
As already shown in Ref.~\cite{Buballa,KKKN}, 
the $\omega^2$ interaction tends to change 
the transition from a first order to a crossover; actually this is seen 
in the result of the NJL+$\omega^2$ model of Fig. 2(a). 
As an interesting result, however, the nonlinear $\sigma^4$ and 
$\sigma^2\omega^2$ interactions make the transition sharper again, so that 
the transition returns to a first order 
in the full-fledged NJL+$\sigma^4$+$\omega^2$+$\sigma^2\omega^2$ model. 
Thus, the nonlinear $\sigma^4$ and $\sigma^2\omega^2$ 
interactions work so as to cancel out the well-known effect of the  
$\omega^2$ interaction.

Figure 3 shows the effective couplings as functions of $\mu$ 
for the case of $T=0$. 
For all models, both $G_{{\rm s}\sigma}^*$ and $G_{{\rm v}\omega}^*$ 
are suppressed in the high $\mu$ region, but each model has its 
own $\mu$ dependence. 
In the NJL+$\sigma^4$ model, 
$G_{{\rm s}\sigma}^*$ decreases suddenly in the high $\mu$ region 
and then approaches the value of $2g_{2,0}$, 
because the $\sigma$-dependent part of $G_{{\rm s}\sigma}^*$ almost vanishes 
there. 
Similarly, in the NJL+$\sigma^4$+$\omega^2$ model, $G_{{\rm s}\sigma}^*$ 
decreases rather rapidly but not suddenly as $\mu$ increases 
and approaches the value of $2g_{2,0}$. 
The change in the $\mu$ dependence of $G_{{\rm s}\sigma}^*$ 
from the the NJL+$\sigma^4$ model to the NJL+$\sigma^4$+$\omega^2$ model 
comes from the fact 
that the $\omega^2$ interaction makes the phase transition weak. 
In the NJL+$\omega^2$+$\sigma^2\omega^2$ model, $G_{{\rm s}\sigma}^*$ 
decreases gradually as $\mu$ increases, 
because the $\omega$-dependent part of $G_{{\rm s}\sigma}^*$ 
gives a negative contribution to the effective coupling. 
In the full-fledged NJL+$\sigma^4$+$\omega^2$+$\sigma^2\omega^2$ model, 
$G_{{\rm s}\sigma}^*$ is suppressed suddenly by 
the $\sigma^4$ interaction and then suppressed gradually 
by the $\sigma^2\omega^2$ mixing interaction.

As for $G_{{\rm v}\omega}^*$ shown in Fig. 3(b), 
one can see a similar sharp suppression 
but not find any gradual decrease. 
This is understandable from the fact that 
in Eq. (\ref{eq:vector-G}) the coupling 
has a $\sigma$-dependent term but no $\omega$-dependent one. 
As a point to be noted, the sharp suppression comes from 
the $\sigma^2\omega^2$ interaction 
for the case of $G_{{\rm v}\omega}^*$ 
but the $\sigma^4$ interaction for the case of $G_{{\rm s}\sigma}^*$. 
Furthermore, note that the $\sigma^2\omega^2$ interaction yields both 
a gradual decrease of $G_{{\rm s}\sigma}^*$ and a sharp suppression of 
$G_{{\rm v}\omega}^*$.

The $\mu$ dependences of the effective couplings mentioned above 
are similar to that of the chiral symmetry restoration shown in Fig. 2. 
We can then consider that effects of higher-order interactions on the chiral 
symmetry restoration are described mainly through the effective couplings 
in their $\mu$ dependences, although 
$U$ is also changed by the higher-order interactions. 

%%%%%%%%%%%%%%%% Fig %%%%%%%%%%%%%%%%%%%%% ->
\begin{figure}[htbp]%[H]
\begin{center}
 \includegraphics[width=7.5cm]{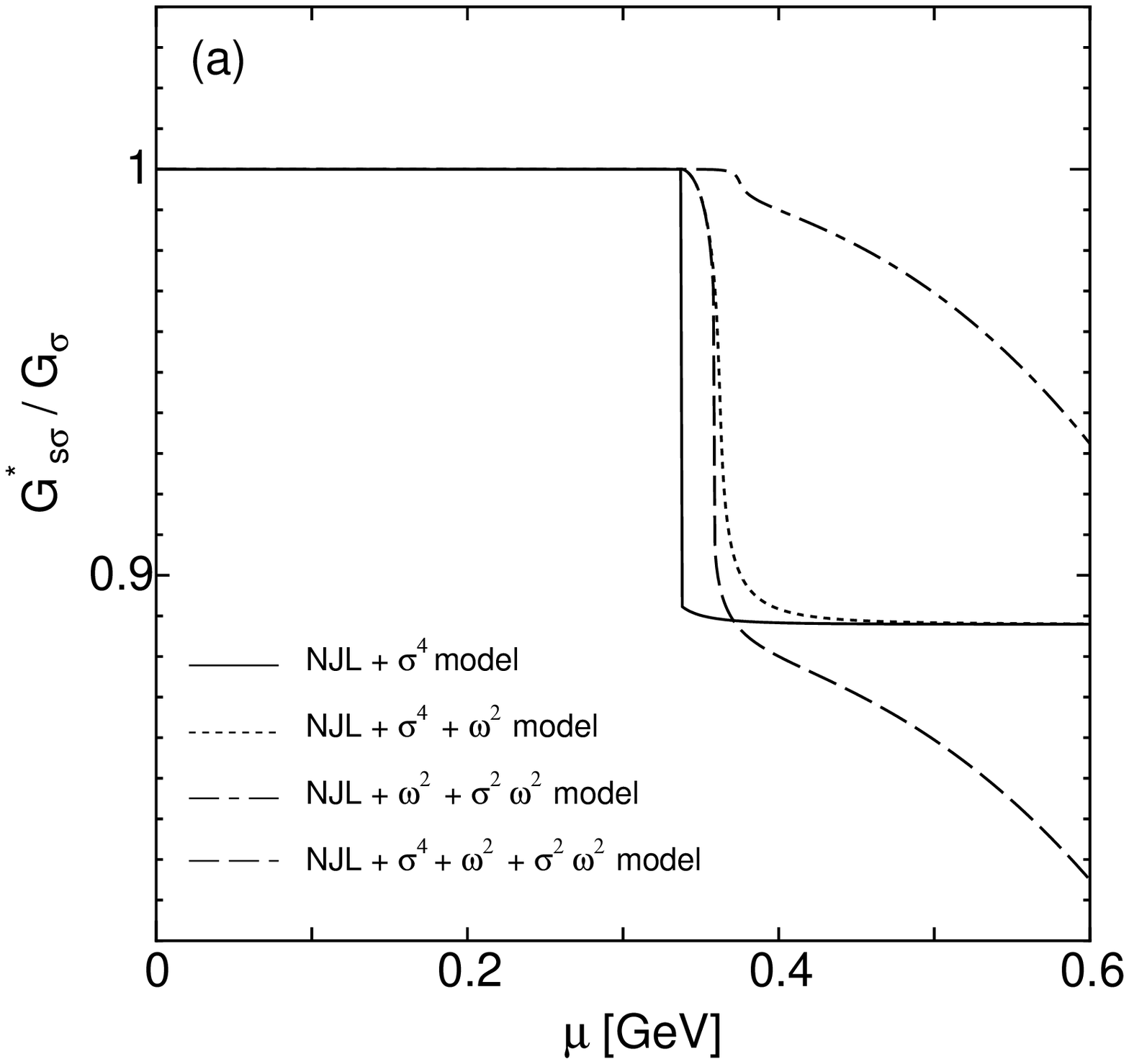} 
 \includegraphics[width=7.5cm]{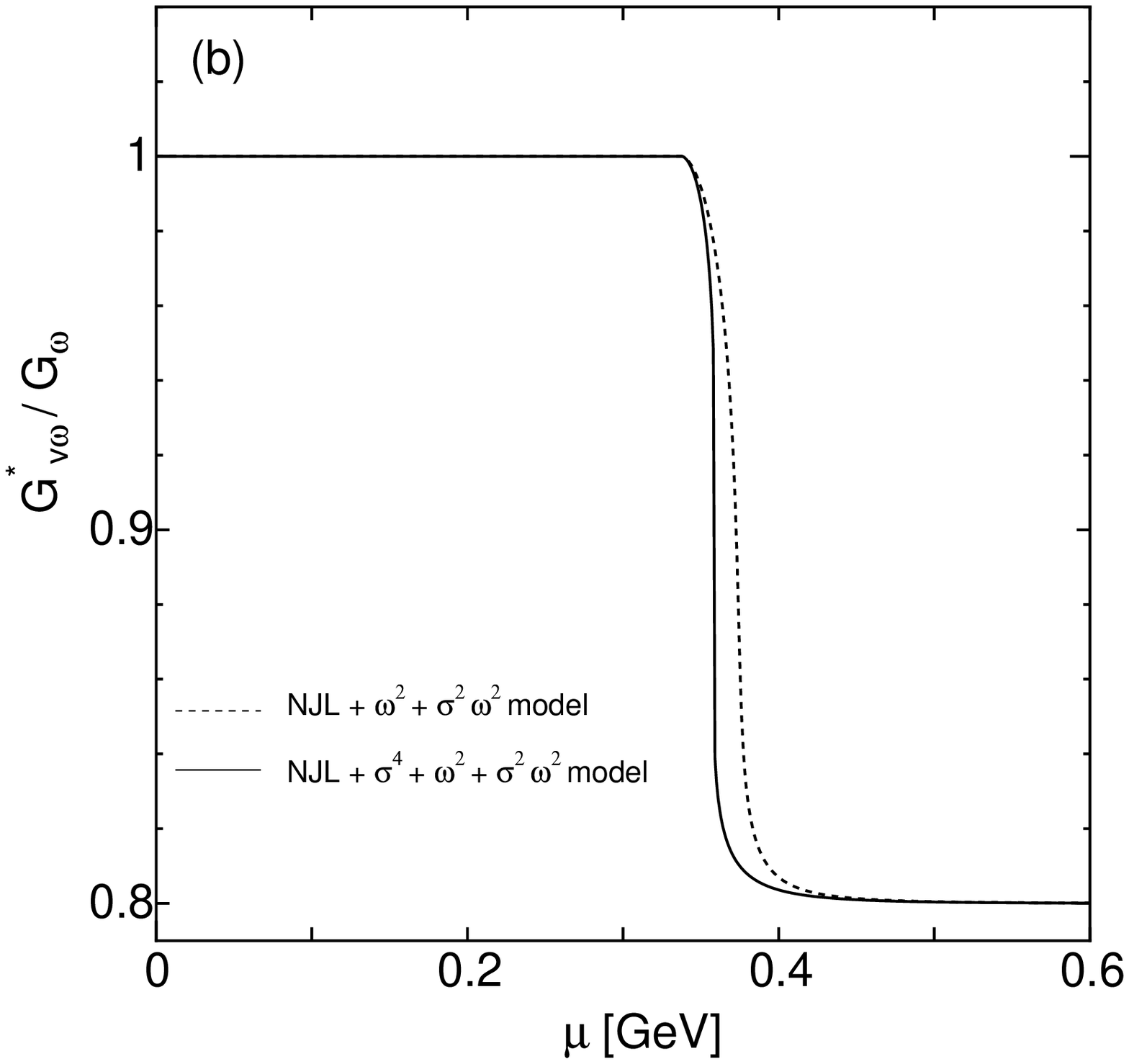}
\end{center}
\caption{Effective couplings as functions of $\mu$ 
in the case of $T=0$. 
}
\label{fig3}
\end{figure}
%%%%%%%%%%%%%%%%%%%%%%%%%%%%%%%%%%%%%%%%%%%%%%%%%

%%%%%%%%%%%%%%%% Fig %%%%%%%%%%%%%%%%%%%%%
\begin{figure}[htbp]%[H]
\begin{center}
 \includegraphics[width=7.5cm]{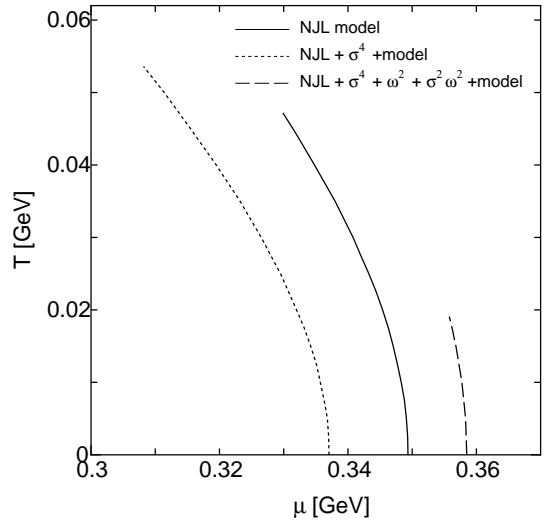} 
\end{center}
\caption{Phase diagram in the $\mu$-$T$ plane. 
Each curve denotes the location of the first-order phase transition in 
each model. 
}
\label{fig4}
\end{figure}
%%%%%%%%%%%%%%%%%%%%%%%%%%%%%%%%%%%%%%%%%%%%%%%%%

Figure 4 shows the phase diagram in the $\mu$-$T$ plane. 
Results are shown for the three cases of the original NJL model, the NJL+$\sigma^4$ model and the NJL+$\sigma^4$+$\omega^2$+$\sigma^2\omega^2$ model 
in which the transitions are first order in the 
high $\mu$ and low $T$ region. 
First-order transitions take place on the curves in Fig. 4, and 
at the endpoint $(\mu_{\rm e},T_{\rm e})$ of each curve the transition is 
changed into a crossover. 
The NJL+$\sigma^4$  model 
yields smaller $\mu_{\rm e}$ and larger $T_{\rm e}$ than 
the original NJL model does; 
$(\mu_{\rm e},T_{\rm e})=(308, 54)$ MeV for the former and 
$(330, 47)$ MeV for the latter. 
When we take another parameter set, 
$\Lambda =0.6315$ GeV, $g_{2,0}=5.00$ GeV$^{-2}$ 
and $g_{4,0}=271$ GeV$^{-8}$, 
that reproduces $M_\sigma =600$ MeV, the endpoint of the NJL+$\sigma^4$ model 
is at $(276, 62)$ MeV. 
Thus, the nonlinear $\sigma$ interaction shifts the critical endpoint 
toward values predicted 
by the lattice QCD calculations, $(\mu_{\rm e},T_{\rm e})=(242,160)$ MeV~\cite{ZF},  
and by the QCD-like theory, 
$(\mu_{\rm e},T_{\rm e})\sim (200,100)$ MeV~\cite{KMT,HTF}.

As mentioned above, the $\omega^2$ interaction tends to change the transition 
from a first order to a crossover, but this effect is partially 
canceled out by the scalar-vector mixing interaction $\sigma^2\omega^2$. 
Consequently, as shown in Fig. 4, there exists a critical endpoint also 
in the NJL+$\sigma^4$+$\omega^2$+$\sigma^2\omega^2$ model. 
The comparison of our models with the results of the lattice simulations 
and the QCD-like theory indicates that the NJL+$\sigma^4$ model is most 
preferable among our models. This may imply that the cancellation 
is essential.

%%%%%%%%%%%%%%%%%%%%%%%%%%%%%%%%%%%%%%%%%%%%%%%%%%%%%%%%%%%%%%%%%%%%%%%%%%%%%%%%%%%%% Summary
%%%%%%%%%%%%%%%%%%%%%%%%%%%%%%%%%%%%%%%%%%%%%%%%%%%%%%%%%%%%%%%%%%%%%%%%%%%%%%%%
In summary, we have studied effects of 
eight-quark interactions on the chiral phase transition. 
The scalar-type nonlinear interaction $\sigma^4$ 
hastens the restoration of chiral symmetry and shifts 
the critical endpoint toward the values predicted by 
the lattice simulations and the QCD-like theory. 
The vector-scalar mixing interaction $\sigma^2\omega^2$  
can make the transition sharper in the high $\mu$ and low $T$ region, 
while the $\omega^2$ interaction works in the opposite direction. 
Thus, the effect of the mixing interaction tends to cancel out that 
of the $\omega^2$ interaction in the high $\mu$ region, 
while the $\omega^2$ interaction is still strong in the normal 
nuclear density region. 
The roles of the eight-quark interactions are well understood 
through the effective couplings, 
$G_{{\rm s}\sigma}^*$, $G_{{\rm v}\sigma}^*$, $G_{{\rm s}\omega}^*$ and 
$G_{{\rm v}\omega}^*$, 
in their $\mu$ dependences. 
Our results indicate that eight-quark interactions 
are very important for phase transitions and must be studied in detail. 
It is very interesting to study effects of the interactions on 
a color superconductivity itself and its correlation with the chiral 
phase transition. The study is now in progress. 
%\bigskip

\noindent
\begin{acknowledgments}
The authors thank M. Tachibana for useful discussions and suggestions. 
H.K. also thanks T. Kunihiro for useful discussions and suggestions. 
\end{acknowledgments}

%%%%%%%%%%%%%%%%%%%%%%%%%%%%%%%%%%%%%%%%%%%%%%%%%%%%%%%%%%%%%%%%%%%%%%%%%%%%%%%%%%%%% References 
%%%%%%%%%%%%%%%%%%%%%%%%%%%%%%%%%%%%%%%%%%%%%%%%%%%%%%%%%%%%%%%%%%%%%%%%%%%%%%%%

\end{document}